# Composition-tuned magneto-optical Kerr effect in $L1_0$-Mn$_x$Ga films with giant perpendicular anisotropy

L. J. Zhu[1,2]*, L. Brandt[1], J. H. Zhao[2], G. Woltersdorf[1]

*1. Institut für Physik, Martin-Luther-Universität Halle, von-Danckelmann-Platz 3, Halle 06120, Germany*
*2. State Key Laboratory of Superlattices and Microstructures, Institute of Semiconductors,*
*Chinese Academy of Sciences, P. O. Box 912, Beijing 100083, China*

**Abstract:** We report the large polar magnetooptical Kerr effect in $L1_0$-Mn$_x$Ga ($0.76 \leq x \leq 1.29$) epitaxial films with giant perpendicular magnetic anisotropy. The Kerr rotation is enhanced by a factor of up to 2.5 by decreasing Mn atomic concentration, which most likely arises from the variation of the effective spin-orbit coupling strength, compensation effect of magnetic moments at different Mn atom sites, and overall strain. A significant tuning effect of composition is also observed on Kerr ellipticity and complex Kerr angle (including the magnitude and phase angle). The good epitaxial compatibility with semiconductors, moderate coercivity of 4.6-9.7 kOe, large Kerr rotation of up to 0.10°, high reflectivity of 35%-55% in a wide wavelength range of 400~850 nm, and giant magnetic anisotropic field of up to 140 kOe together make these $L1_0$-Mn$_x$Ga films promising for scientific and technological applications in spintronics and terahertz-frequency magnetooptical modulators.
* Author to whom any correspondence should be addressed. Email: zhulijun0@gmail.com

## 1. Introduction

The polar magneto-optical (MO) Kerr effect is a powerful tool that can locally probe the magnetic properties and electronic structures of materials [1] and modulate the optical polarization at the spin precession frequency [2]. Most recently, a variety of new exciting spintronic phenomena have been observed taking advantage of the polar MO Kerr effect, such as voltage-controlled magnetic anisotropy in ferromagnetic thin films [3], broken time-reversal symmetry in the heavy-fermion superconductors and bilayer graphene [4,5], terahertz spin precession in perpendicularly magnetized Mn₃Ge films [6], spin Hall effect in semiconductors [7], magnetic skyrmion bubbles [8], magnetic vortex dynamics [9], spin orbit torques [10], and nanosecond current-induced domain wall motion [11] in ferromagnet/heavy metal bilayers and nanowires. From the viewpoint of technological application, high-speed optic communication requires modulators of light magnitude/polarization working at very high frequencies. For MO modulators, the conventional ferromagnetic materials, e.g. Permalloy, usually have very weak magnetic anisotropy which limits the modulation bandwith to a few GHz or less [2]. MO materials with giant magnetic anisotropic field ($H_k$) of ~100 kOe, e.g. Mn₃Ge [6], are good candidates for ultrafast MO modulators due to their terahertz-frequency spin precession, which is at least two orders of magnitude faster than that based on conventional ferromagnetic materials [2]. Furthermore, MO materials epitaxially compatible with semiconductors allow for direct integration of MO functional devices with underlying photonic circuits. Therefore, for both the spintronic and modulator applications [6-11], it is highly desirable to develop new kinds of MO materials which simultaneously have large saturation Kerr rotation ($\theta_K$), high optical reflectivity ($R$), giant $H_k$, and good semiconductor compatibility, and to further tailor their MO properties.

In past two decades, the noble-metal-free and rare-earth-free Mn$_x$Ga alloys with giant uniaxial magnetic anisotropy ($K_u$~20 Merg/cc), low Gilbert damping constant, and good semiconductor compatibility have attracted increasing attention for their great potential in ultrahigh-density perpendicular magnetic recording, spintronics, and permanent magnet applications [12-14]. The long-range chemical ordering of Mn$_x$Ga alloys is well-known to be $L1_0$ for $0.76 \leq x \leq 1.5$ and $D0_{22}$ for $1.8 \leq x \leq 3$, respectively [12,13]. There is also a wide consensus about their magnetic properties that varies profoundly with composition [15,16], crystalline quality, and chemical ordering degree [17]. However, the understanding of the MO properties of Mn$_x$Ga has been remained incomplete. Krishnan found a large $\theta_K$ of 0.1° at wavelength $\lambda$ = 820 nm and high $R$ in a 30 nm-thick Mn$_{1.5}$Ga film and pointed out a promise for MO recording application [18]. In contrast, a negligible $\theta_K$ (0.01° at $\lambda$ = 633 nm) was reported later in a Mn₂Ga bulk sample [19], making the application promise problematic. The violating MO performance of the same material system can be due to a difference in the structural quality and composition of their samples. It should be also noticed that a comprehensive understanding of MO properties of a material needs knowledge from both Kerr rotation and ellipticity ($\varepsilon_K$), or the magnitude ($|\psi|$) and phase angle ($\alpha$) of the complex Kerr angle $\psi = |\psi|e^{i\alpha} = \theta_K + i\,\varepsilon_K$. Unfortunately, there have been, up to date, no report about $\varepsilon_K$ and $\psi$ for this material.

On the other hand, MO samples with high structural quality and good magnetic performance are desirable for their potential scientific and technological applications in spintronics and MO modulators. However, high-quality Mn$_x$Ga epitaxial films with large $x$ ($1.5 \leq x \leq 3$) always show huge magnetization switching field ($H_c$) (e.g. 4.3 T for $x$ = 1.5) and very small magnetization ($M_s$) [15-17], which substantially limit their practical applications. In our previous work, we found that tuning down the Mn/Ga atomic ratio ($x$) is a simple but effective way to cut down $H_c$ and to enhance $M_s$. Importantly, the high-quality crystalline structure as well as the giant $H_k$ can survive even if $x$ goes down to 0.76 [16]. However, the MO properties of $L1_0$-Mn$_x$Ga films in the low Mn concentration regime ($x<1.5$) remains unknown.

In this paper, we, for the first time, systematically studied the polar MO Kerr effect in $L1_0$-Mn$_x$Ga films in the small-$x$ regime ($0.76 \leq x \leq 1.29$). We found that the magnitude and phase angle of the complex Kerr angle $\psi$ (see Fig. 1(a)) can be effectively tailored by varying the composition, while the reflectivity appears almost independent of $x$. The combination of large $\theta_K$ of up to 0.10°, high $R$ of 35%~55%, and giant $H_k$ of up to ~140 kOe makes these small-$x$ $L1_0$-Mn$_x$Ga films good candidates for both MO detection of the emerging spintronic phenomena and terahertz MO modulators.

## 2. Experimental methods

A series of $L1_0$-Mn$_x$Ga single-crystalline films with different $x$ were deposited on 150 nm-GaAs-buffered semi-insulating GaAs (001) substrates by molecular-beam epitaxy (see Fig. 1(b)) [16]. The growth temperature for $L1_0$-Mn$_x$Ga layers was chosen to be 250 °C, which was found to give the best magnetic performance, the highest chemical ordering, and weakest electron scattering at impurities [17,20]. $x$ was designed by carefully controlling the Mn and Ga fluxes during growth and later verified by x-ray photoelectron spectroscopy to be 0.76, 0.97, 1.13, and 1.29 [16]. After cooling down to room temperature, each film was capped with a 1.5 nm-thick MgO layer to prevent oxidation. The imaginary part of the refractive index of MgO is negligible in the visible light regime [21], such that the measured MO properties are not influenced by the ultrathin capping layers and can be taken as intrinsic to $L1_0$-Mn$_x$Ga layers. As was determined by x-ray reflectivity measurements, the film thickness varies between 42 nm and 49 nm, which is safely larger than penetration depth of the detection light for MO properties (typically ~10 nm). The x-ray diffraction studies indicate that the out-of-plane lattice constant $c$ decreases from 3.644 Å to 3.489 Å with increasing $x$ [16], suggesting a remarkable change in the overall strain. As reported previously, as $x$ increases from 0.76 to 1.23, $K_u$ increases monotonically from 8.6 Merg/cc to 19.3 Merg/cc, while the saturation magnetization ($M_s$) drops from 445 emu/cc to 277 emu/cc. Figure 1(c) shows a schematic of the optical setup used for polar MO Kerr effect measurements. Two laser diodes were used for measurements at $\lambda$=400 nm and 670 nm, respectively. The angle between the incident and reflected beams was fixed at 4°. The magnetic field ($H$) of up to $\pm$ 20 kOe was applied along the sample normal by an electromagnet. The reflectivity measurements were performed at room temperature by a JASCO V-670 spectrophotometer with unpolarized incident light normal to the sample surface and an integrating sphere for collecting the reflected light. In this configuration, the measured reflectivity is not influenced by neither the polarization nor the incidence angle of the light.

## 3. Experimental results and discussions

Figures 2(a) and 2(b) show the hysteresis loops of polar Kerr rotation and ellipticity at $\lambda$ = 400 nm for $L1_0$-Mn$_x$Ga films with different $x$. These well-defined hysteresis loops indicate strong perpendicular anisotropy in these films. The coercivities for rotation and ellipticity hysteresis consistently increase from 4.6 kOe to 9.7 kOe as $x$ increases, which is much smaller than that of high Mn-concentration $L1_0$- or $D0_{22}$-ordered Mn$_x$Ga ($x > 1.5$) epitaxial films grown on Cr-MgO [15] or GaAs [16]. The moderate $H_c$ of these small-$x$ films makes them highly immune to external magnetic noises but easily switchable with an applied magnetic field. Notably, the Kerr rotation and ellipticity show different signs for each film, i.e. $\theta_K >$ 0, while $\varepsilon_K < 0$ at positive saturation states, suggesting that $\alpha$ lies between $-\pi/2$ to 0 as discussed later. The same features hold for the hysteresis loops at $\lambda$ = 670 nm.

Figures 3(a) summarizes $\theta_K$ in $L1_0$-Mn$_x$Ga films as a function of $x$. $\theta_K$ increases quickly from 0.06° (0.04°) for 400 (670) nm at $x$=1.29 to a maximum value of ~0.10° for both wavelengths as $x$ is tuned down to 0.76. The large value of $\theta_K$ for $L1_0$-Mn$_{0.76}$Ga proves the great promise of this material for MO applications. Microscopically, the MO Kerr effect arises from the spin-orbit interaction and exchange splitting which is a function of $M_s$. *Ab initio* and relativistic band-structure calculations show that $\theta_K$ scales linearly with spin-orbit interaction strength ($\xi$) [22-25]. The large $\xi$ was directly responsible for large Kerr effect in alloys with 4$f$ rare-earth elements or 4$d$ Pt, e.g. CeSb [26], MnBi [27], and FePt [3]. As for Mn$_x$Ga films, Mn and Ga are light elements with $\xi$ of ~40 meV (3$d$) and ~155 meV (4$p$), respectively [28]. The small $\xi$ of $L1_0$-Mn$_x$Ga was also considered responsible for the low Gilbert damping constant in pump-probe MO Kerr effect measurements [14]. As indicated by recent experimental and *ab initio* studies on the anomalous Hall effect [29], Gilbert damping [30], and the polar MO Kerr effect [31], the effective values of $\xi$ in alloy films can be continuously tuned by varying the composition. Since the spin-orbit interaction of the Mn 3$d$ state is almost 4 times weaker than that of the Ga 4$p$ state, the decreasing $\theta_K$ with increasing $x$ can be attributed to the decreasing effective $\xi$. Here the average value of $\xi$ of Mn$_x$Ga drops from 105.3 meV to 90.9 meV as $x$ increases from 0.76 to 1.29 (see Fig. 3(b)), if $\xi$ of each element is independent of the composition [28]. On the other hand, $M_s$ decreases monotonically as $x$ increases from 0.76 to 1.29 [16]. First principles calculations suggest strong hybridization of Mn 3$d$ states with Ga 4$p$ states near the Fermi level through spin-orbit interaction [32]. The density of states around Fermi level was found to be significantly suppressed with decreasing $c$ as a consequence of the increase of the interatomic hybridization. The suppression of the density of states will weaken the tendency towards ferromagnetism according to Stoner criterion for itinerant ferromagnetism [33]. Accordingly, the exchange splitting and $M_s$ also decrease with decreasing $c$ value [16]. In the chemically disordered Mn$_x$Ga alloys, a portion of the Mn atoms are expected to antiferromagnetically couple to the other moments [15], leading to suppression of the exchange splitting and compensation of $M_s$. The compensation effect can be inferred to be enhanced with $x$ increasing toward 1.29, although an accurate quantitative determination is difficult. Therefore, the $x$-dependence of the Kerr rotation can be partly attributed to the variation of the exchange splitting as a consequence of the shrinking $c$-lattice axis (see Fig. 3(b)) and the increasing compensation effect of Mn moments at different sites with increasing Mn concentration. Here, it should be pointed out that the thickness of these films is safely larger than the penetration depth of the detection light, and therefore not responsible for the variation of $\theta_K$. It should also be noticed that the magnetic anisotropy is not expected to exert any influence on the MO properties [24]. Based on above discussions, the previously reported large $\theta_K$ in a $Mn_{1.5}Ga$ film (0.10° at 820 nm) [18] in comparison to our samples should be mainly attributed to its longer $c$-axis (3.515 Å) and higher $M_s$ (460 emu/cc). The small $\theta_K$ of 0.01° reported in the $Mn_2Ga$ bulk [19] may be due to the enhanced compensation effect and possible poor crystalline quality in comparison to our epitaxial films. It should be pointed out that $M_s$ can be significantly improved by post-annealing in vacuum [16], which provides possibility to further enhance the Kerr rotation of our $Mn_xGa$ films.

Figures 3(c)-3(e) show $\varepsilon_K$, $|\psi|$, and $\alpha$ for the $L1_0$-$Mn_xGa$ films with different $x$. With increasing $x$, $\varepsilon_K$ varies monotonically from -0.12° to -0.10° for $\lambda = 400$ nm, while changes from -0.04° to -0.07° for $\lambda = 670$ nm, respectively. The opposite dependences of $\varepsilon_K$ on $x$ suggest that $\varepsilon_K$ does not necessarily scale linearly with the magnitude of $M_s$, $\xi$, or $\theta_K$ for samples with different composition, although Kerr ellipticity in magnetic materials is expected to be closely related to Kerr rotation [23]. Theoretical calculations taking into account all different variables (composition, spin-orbit interaction strength, strain, compensation effect, and different electronic transitions [23,24]) would be helpful for better understanding the $x$-dependence of $\varepsilon_K$ in the $L1_0$-$Mn_xGa$ films. For both 400 nm and 670 nm, $|\psi|$ and $\alpha$ decreases monotonically with increasing $x$, indicating a significant tailoring effect of the composition on the complex Kerr angles of the $L1_0$-$Mn_xGa$ films.

Figure 4(a) shows the room temperature optical reflectivity spectrum of the $L1_0$-$Mn_xGa$ films with different $x$. All of these films show high $R$ of 35%~55% in the wide wavelength range from 400 nm to 850 nm. The high reflectivity of these films is highly favorable for both scientific and technical applications of their polar MO Kerr effect.

It is also worth mentioning that, besides the large Kerr rotation and high reflectivity, a giant internal magnetic anisotropic field is crucial for high-frequency MO modulator application. In single-crystalline $L1_0$-$Mn_xGa$, $H_K$ can be determined by $H_K = 2K_u/M_s$, where $K_u$ is the uniaxial magnetocrystalline anisotropy constant deduced from the area enclosed between the magnetization curves in applied fields parallel and perpendicular to the film plane [17]. The spin precession frequency ($f$) due to $H_K$ can be estimated at the first rough approximation following $f = g\mu_B H_k/h$, where $g$, $\mu_B$, and $h$ are the $g$-factor (~2.0), Bohr magnetron, and Plank constant, respectively. As is shown in Fig. 4(b), $H_k$ increases from ~39 kOe to ~140 kOe with increasing $x$. Similar to the case of $Mn_3Ge$ [6], the square perpendicular magnetization hysteresis and the giant $H_k$ should give rise to a spin precession at 0.2-0.4 terahertz frequency, allowing for ultrafast optical modulation. Brillouin light scattering and pump-probe MO Kerr effect experiments would be needed in the future to obtain detailed data on the terahertz-frequency spin precession expected in this material.

## 3. Conclusion

We have presented the composition effects on the polar MO Kerr effect in $L1_0$-$Mn_xGa$ ($0.76 \leq x \leq 1.29$) epitaxial films with giant $H_k$. The Kerr rotation is enhanced from 0.04° (0.06°) to 0.10° as $x$ is reduced from 1.29 to 0.76 for wavelength of 400 (670) nm, which is attributed to the variation of effective spin-orbit coupling strength, compensation effect of magnetic moments at different Mn sites, and overall strains. A significant composition-tuning effect was observed on the magnitude and the phase angle of the complex Kerr angle. These films also show reflectivity of 35%~55% in a wide wavelength range of 400~850 nm. The combination of the good semiconductor compatibility, the moderate coercivity, the large polar MO Kerr effect, the high $R$, and the giant $H_k$ makes these small-$x$ $L1_0$-$Mn_xGa$ films a very interesting MO material. Our results would be much helpful for understanding of MO properties of $Mn_xGa$ in the low Mn concentration regime and approaching the practical applications of this material in the high-sensitivity and high magnetic-noise-immunity MO studies of the emerging spintronic phenomena and in terahertz-frequency MO modulator applications for ultrafast optic communications.

## Acknowledgments

A fruitful discussion with P. M. Oppeneer is gratefully acknowledged. This work was supported by priority program (SPP 1538) of the German research foundation (DFG), NSFC (Grand No. 61334006), and MOST of China (Grant No. 2015CB921503).

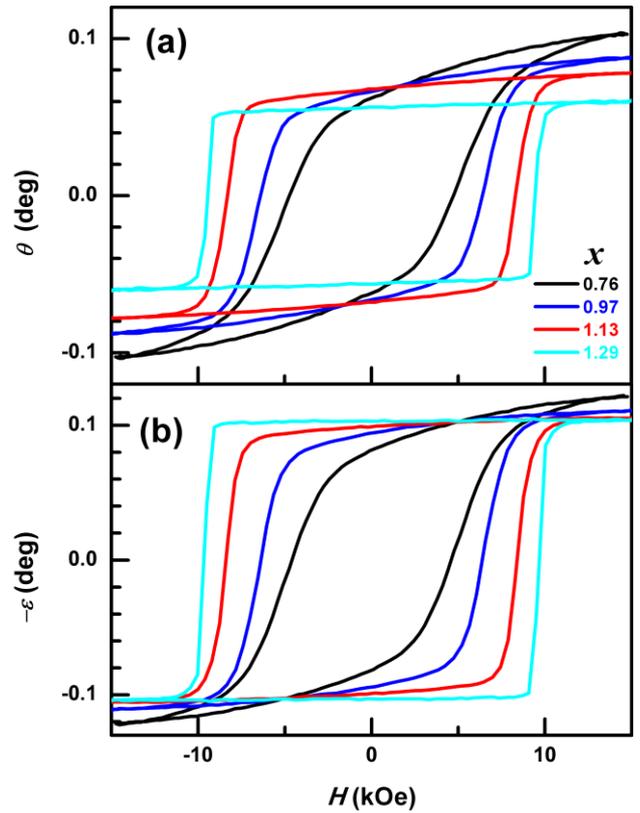

Figure 2. Hysteresis loops of polar Kerr (a) rotation ($\theta$) and (b) ellipticity (-$\varepsilon$) at 400 nm for $L1_0$-Mn$_x$Ga films with different $x$.

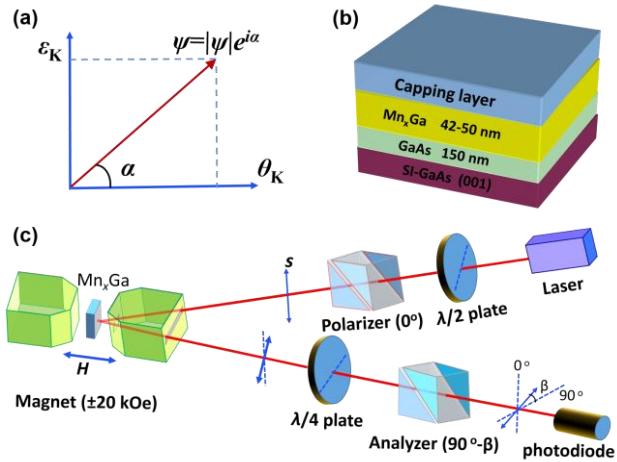

Figure 1. Schematic of (a) complex Kerr angle, (b) sample structure, and (c) polar MO Kerr effect setup. $\beta$ represents the bias angle of analyzer polarization away from the full extinction direction.

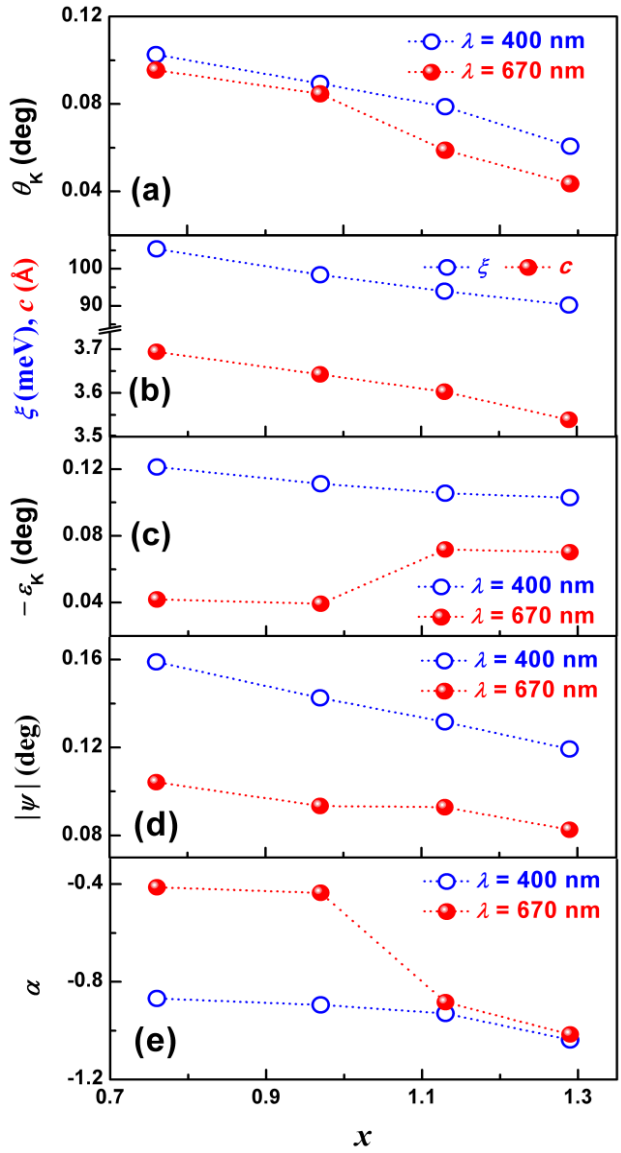

Figure 3. $x$-dependence of (a) $\theta_K$, (b) $-\varepsilon_K$, (c) $c$ and $\xi$, (d) $|\psi|$, and (e) $\alpha$ for $L1_0$-Mn$_x$Ga films. The data for $c$ was taken from [16].

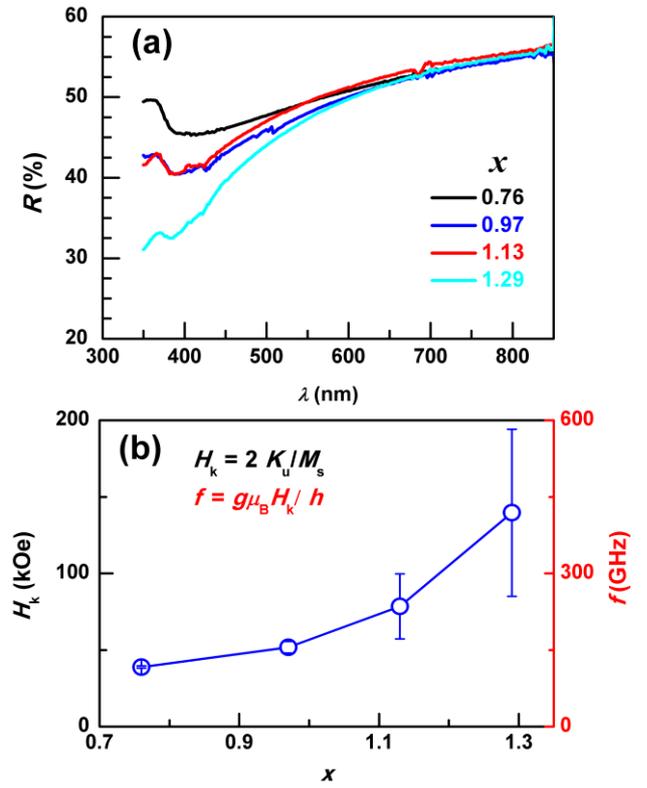

Figure 4. (a) Reflectivity spectrum and (b) the perpendicular anisotropic fields ($H_k$) and the estimated spin precession frequencies ($f$) of $L1_0$-Mn$_x$Ga films with different $x$.